# Scanned SQUID Microscope with High-speed Electrical Connectivity

Ian W. Haygood[1], Bochao Xu[2,3], John Biesecker[1], Michael L. Schneider[1]

[1] National Institute of Standards and Technology, Boulder, Colorado 80305, USA
[2] Associate of the National Institute of Standards and Technology, Boulder, Colorado 80305, USA
[3] Department of Physics, University of Colorado Denver, Denver, Colorado 80217, USA

We report on a scanned superconducting quantum interference device (SQUID) microscope operating in a cryogen-free cryostat, with the capability of up to forty RF connections with 20 GHz bandwidth to a device under test. The system utilizes planar gradiometric DC SQUIDs, which are fully shielded except for a pair of pickup coils with radii as small as 250 nm and have on chip field coils allowing for susceptometry. The system noise is 1.3 $\mu\Phi_0/\sqrt{Hz}$ at the base temperature of 3.3 K. The sample temperature is variable and both magnetometry and susceptibility measurements are simultaneously possible with the sample temperature above 40 K. Through the use of a cryogenic chip socket and silicon interposer round trip RF losses to a sample are approximately 15 dB at 20 GHz. A combination of both active and passive magnetic shielding result in a residual magnetic field less than 100 nT at the sample location.

The use of a superconducting quantum Interference device (SQUID) as a sensor for scanned probe microscopy was originally demonstrated in 1983[1], where a 230 µm diameter coil, inductively coupled to an RF SQUID, was linearly scanned across a high-$T_c$ superconducting film. Two-dimensional scanning was later demonstrated with a 1 mm diameter pickup coil and vortices were imaged in a niobium film.[2] The resolution of the two-dimensional systems were significantly improved by the adoption of lithographically defined DC SQUIDs.[3,4] Further improvements to resolution and flux sensitivity were made by going to smaller pickup loops and by galvanically[5–7] coupling to the SQUID rather than using a flux transformer. The utility of scanned SQUID sensors was further increased by adding a counter-wound pickup loop, resulting in a gradiometric design, and by adding field coils, allowing for local susceptibility measurements to be performed.[8,9] Scanned vector magnetometers were also demonstrated.[10] Submicrometer spatial resolution has been achieved using both planar gradiometric susceptometer designs[11] and SQUID on a tip (SOT) sensors.[12–17] Recently gradiometric susceptometers fabricated using nano-constrictions have been demonstrated which have low flux noise and can operate in magnetic fields up to 1.4 T.[18–20]

Scanned SQUID microscopes (SSM) utilizing the previously mentioned sensors have high magnetic sensitivity, high special resolution, and low flux noise making them good platforms for studying magnetic properties[21–23], current distributions[24,25], and superconducting film properties[26]. For example, the effectiveness of moat structures, designed to intentionally trap flux in specific regions in superconducting digital logic circuits can be studied[27,28].

Finally, superconducting device properties such as the current-phase relationship in Josephson Junctions (JJs) can be measured using SSMs[29,30].

Historically these systems have been designed to operate in a liquid helium dewar due to the high cooling power, low vibration, and simplicity of the designs. More recently, due to high cost and limited availability of liquid helium, several cryogen free systems have been demonstrated[31–33]. Such systems can have a low operating cost and stay cold for long periods of time, but they have the disadvantage of increased mechanical vibrations and electrical noise introduced by the cryo-cooler.

In this paper we introduce an SSM designed around a Gifford-McMahon (GM) cooler which sits to the side of an optical table, onto which the cold space is bolted. By moving the cryocooler off of the optical table and onto the ground, system vibrations are minimized. Additionally, the system has the capability to have up to 40 RF coaxial lines, and more than 50 DC lines running from room temperature to the 4 K plate while still maintaining a base temperature of 3.3 K. The system can also incorporate a cryogenic interposer, described below, which maintains open access to a device under test (DUT) for the SSM sensor. Together the system with the interposer allows broadband high-speed signals to reach the sample with low loss without disrupting SQUID magnetometer access. We demonstrate imaging of vortices in a niobium film, susceptibility measurements at variable sample temperatures, and imaging of magnetic fields from current lines. Finally, we discuss the goal of using this system to measure electrical properties and operating margins in superconducting digital logic circuits and SSM measurements in a single cooldown, allowing for direct correlation between the presence and location of flux vortices and electrical circuit performance.

**SYSTEM OVERVIEW**

The SSM is built upon a Four Nine Design "SideKick SK 300" cryostat[34]. The cooling is provided by a 1 W Gifford-McMahon (GM) cryocooler which sits in a carriage system with differentially pumped bellows, removing the need for springs to oppose the pressure differential in the vacuum space. Additionally, the carriage system effectively shunts the mechanical energy from the GM piston displacement to the laboratory floor. Furter reductions in transmitted mechanical energy are achieved by connecting the first and second stage of the GM cooler with compliant aluminum thermal straps to the cold spaces, which are bolted to a non-magnetic aluminum optical table with a composite core. An overview showing the external form factor of the system is show in Figure 1(a).

The cold space of the cryostat can be fitted with a variety of sample holders and a configuration with a variable sample temperature mount is shown in Figure 1(b). In order to adequately cool the coaxial lines in the system two towers were installed on the 4 K plate, one of which can be seen on the right side of the picture. The coaxial lines are thermally sunk to the top portion of the tower and the added height allows for the thermal gradient required for 4 K operation. The coarse positioners and scanning stages are mounted to a copper roof and in this system the sensor is scanned while the sample is fixed. Mounted to the top of the scanner stages is a printed circuit board (PCB), shown in Figure 1(c). Soldered to the top of the PCB is a 75 µm thick cantilever attached to a 375 µm stand off from a bottom electrode. The PCB provides electrical connectivity for both the SQUID sensor and for sensor contact detection. This is done by measuring the capacitance of the cantilever using a modulated supply voltage amplified by a charge amplifier. The resulting signal using this method is around 325 µV/µm of displacement. Using a lock-in measurement we can measure changes on the order of 1 µV resulting in quick, reliable, and low-force contact detection.

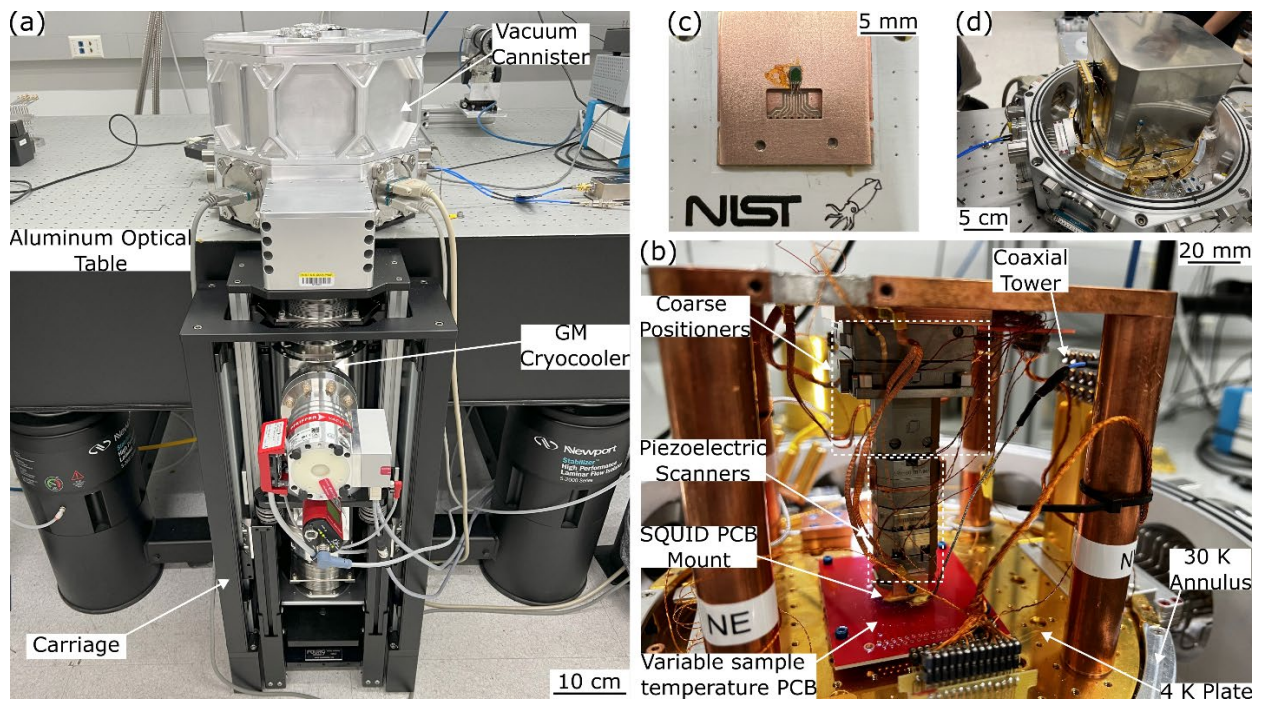

Figure 1: Optical images of various parts of the SSM system. (a) Overview showing the cold space bolted to the optical table and the vacuum bellows pumping on both sides of the GM cooler in the carriage system. (b) a closeup of the cantilever PCB with a susceptometer wire bonded. (c) 4 K mu-metal passive magnetic shielding. (d) Image of the 200 mm 4 K stage surrounded by the 30 K annulus. The coarse and fine scanners are mounted to a copper roof and the DUT is located on a variable sample temperature PCB thermally anchored to the 4 K stage.

Figure1(d) is a photo of the passive mu-metal shielding installed in the system. To fit the shielding in as close as possible without modifying the 30K radiation shield and vacuum canister the magnetic shielding was designed to fully enclose the scanners and sample, with only small holes for DC and coaxial lines. Even with a less-than-optimal aspect ratio the passive shielding results in a background field of 1 µT at the sample, and external field coils can bring this down to below 100 nT. The passive shielding also makes the system much less susceptible to environmental noise and locked-loop operation has been maintained for weeks.

The system is configured to bring up to 40 high-speed lines to the sample. It has two ISO 63 flanges, each with 20 hermetic sub miniature push-on (SMP) feedthroughs. From the SMP feedthroughs semi-rigid cables provide signals via heat sinking on the 30 K annulus to SMP bulkheads on towers which are thermally anchored to the 4 K stage. Flexible coaxial lines bring signals from the bulkhead to a printed circuit board designed to transmit high-speed signals via a high-density RF connector routed to a cryogenic chip socket shown in Figure 2(a). The chip socket can be fitted with an interposer, shown in Figure 2(b), which has co-planar waveguides (CPW) to route signals to the sample at the center. High bandwidth electrical and mechanical connection to the sample is done using 5 µm thick indium bump bonds on the interposer inner pads. A through wafer deep reactive ion etch (RIE) is done on the interposer to open an aperture. Figure 2(b) inset, shows a closeup of the other side of the interposer showing the sensor access to the DUT while electrically connected. Initial characterization of the entire RF signal path, Figure 2(c), shows typical round trip losses (one external SMP port to an adjacent SMP port) of around 15 dB at 20 GHz with much of the loss coming from the semi rigid cables and interconnects used in the cryostat. The lower blue curve represents the total loss, including all room temperature and cryogenic cabling. The upper black curve has the external and semirigid cable losses removed by capturing a trace with a 6 inch cable as a thru on the SMP connectors on the tower. This trace was then subtracted to produce the black "To Tower" data.

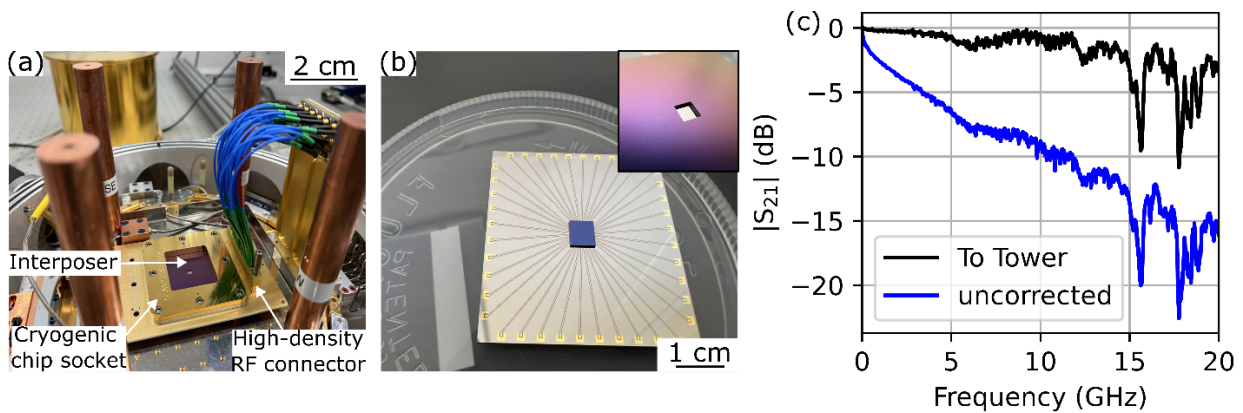

Figure 2: Overview and performance of the high-speed signals interfacing with the interposer with optical access. (a) optical image of the PBC interfacing with the high-density cable assembly connector and chip socket. (b) Close up of the 38 mm interposer with 40 CPW lines leading to the indium bump bonded DUT. The inset shows the other side of the interposer, highlighting access for the sensor. (c) Scattering parameters for a DUT with a CPW connecting adjacent lines.

The gradiometric magnetometer/susceptometer sensor used for imaging in the system is similar to previous designs[9,11] with two significant modifications, which have both benefits and disadvantages. First, the modulation coil and JJs have been moved outside of the coaxial body of the sensor akin to a design from Garner et al.[35] This results in a 200 µm² pickup area which is susceptible to background fields and gradients. While this is a disadvantage compared to the previously mentioned designs it allows easy incorporation of the second modification, a damping resistor between the two legs of the SQUID loop. This damping resistor reduces resonances in the sensor, eliminating "steps" in the current-voltage characteristics and reduces sensor noise[36].

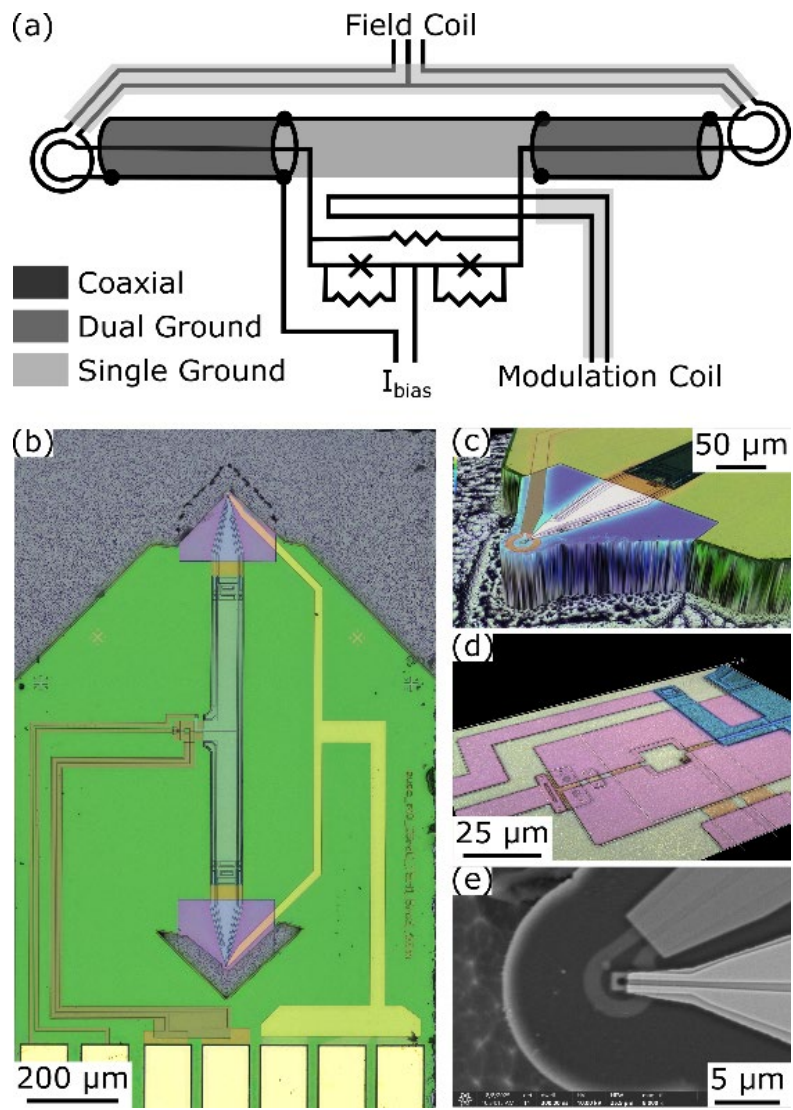

Figure 3: Overview of gradiometric magnetometer/susceptometer. (a) Electrical schematic of the gradiometric susceptometer/magnetometer sensor. (b) Optical micrograph overview of a sensor. (c) Optical profilometer image showing the deep RIE near the pickup coil. (d) Optical profilometer image of the modulation coil. The SQUID JJs, shunt resistors, and damping resistor can all be seen alongside the modulation coil (e) Electron micrograph of a 500 nm diameter pickup loop with 3 µm diameter field coil.

An electrical schematic highlighting the above changes is shown in Figure 3(a). An optical profilometer image showing the external modulation loop along with the damping resistor, shunt resistors, and JJs is shown in Figure 3(d). The sensor is 1600 μm x 900 μm and an overview image is shown Figure 3(b). Around the pickup coil and counter pickup coil, 50 μm deep RIE has been performed, shown Figure 3(c), resulting in a consistent edge to field coil distance, and eliminating the need to hand polish sensors. Sensors with pickup coil radii from 250 nm (Figure3(e)), up to 1.3 μm are available allowing us to choose an ideal sensor for each measurement.

**SYSTEM PERFORMANCE**

**Flux Noise**

The flux noise of the system is an important metric of the overall system performance and sets a lower limit on the sensitivity of both magnetometry and susceptometry measurements. The flux noise is characterized by capturing a 10 second time trace of the voltage response of a sensor in locked-loop operation with the magnetic shielding installed. The voltage power spectral density (PSD), $S_V$, can be calculated and then converted to the flux PSD, $S_\Phi$, using $S_\Phi = S_V/V_\Phi^2$, where $V_\Phi$ = 14.7 V/$\Phi_0$ for our feedback loop operating in high sensitivity. The full spectrum is plotted in Figure 4 and is relatively flat from 10 Hz – 10 kHz with the exception of peaks correlating to the power line frequency of 60 Hz and higher harmonics. There is also 1/f noise starting below 2 Hz. We quantify the flux noise to a single number by averaging the response between 500 Hz – 1 kHz, giving $S_\Phi$ = 1.3 μ$\Phi_0/\sqrt{Hz}$. This low value of flux noise was only achievable by careful shielding of the SQUID series array amplifier, as well as the sample space.

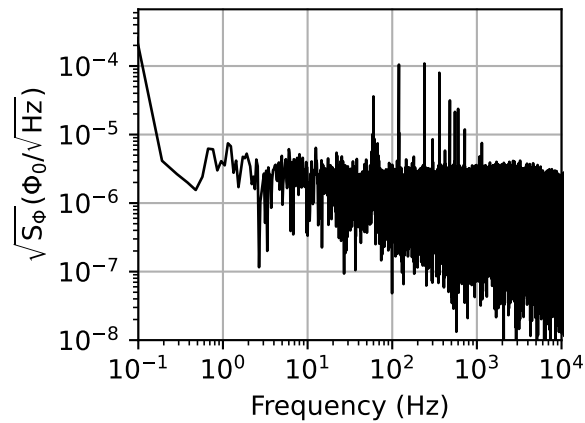

Figure 4:Flux noise power spectral density for the system operating at 3.3 K with magnetic shielding installed. The average flux noise between 500 Hz – 1 kHz is $S_\Phi$ = 1.3 μ$\Phi_0/\sqrt{Hz}$

**Vibrations**

The cryostat which the system was built upon was chosen to effectively minimize the vibrations of the cold plate measured relative to the optical table. To benchmark the in-plane vibrations in the system we used a method described in several other papers[31–33,37] and imaged a single vortex, capturing a one-second time trace at each scanned position in a 58 x 75 pixel (0.03 µm$^2$) array. An image of the vortex, obtained using a 500 nm diameter pickup loop, used to determine the vibrational spectrum is shown in Figure 5(a), where each pixel is the time averaged value from the data capture. The calculated vibrational spectrum for the x (parallel to sensor long axis), and y (perpendicular to sensor long axis) axes are plotted in Figure 5(b). The amplitude of the vibrations is generally lower in the x-axis compared to the y-axis which is unexpected given the relative symmetry of the coarse and fine positioning piezoelectric stacks. The RMS vibrational amplitude is 116 nm in the x-axis and 257 nm in the y-axis which is well below the dimensions of the pickup loops.

The vibrations in the out-of-plane direction (z-axis) were characterized by performing susceptibility measurements on a Nb film. The susceptibility was first measured while approaching the sample and then the sensor was positioned 400 nm above the sample. At this height the response of the susceptibility is approximately linear as a function of separation for the vibrational amplitudes observed with dz/dV = 8.7 µm/V. The field coil was driven at 10 kHz where the flux noise is around 500 nΦ$_0$/Hz$^{1/2}$ resulting in a noise floor for the measurement of 68 pm displacement. To ensure that the higher frequency signals are not significantly attenuated from the lock-in low-pass filter, the lock-in the time constant was set to 100 µs with a 24 dB/octave roll off, resulting in -3 dB attenuation at 692 Hz. The vibrational spectrum obtained from the susceptibility is show in Figure 5(c) with both the compressor running and when the compressor has temporarily been turned off. The out-of-plane vibrations are generally below 1nm/Hz$^{1/2}$ with the exception of a broadband set of peaks around 245 Hz which is present weather or not the cooling system is running. The RMS vibrations are 25 nm with the cooling on and are reduced to 17 nm with the cooling off.

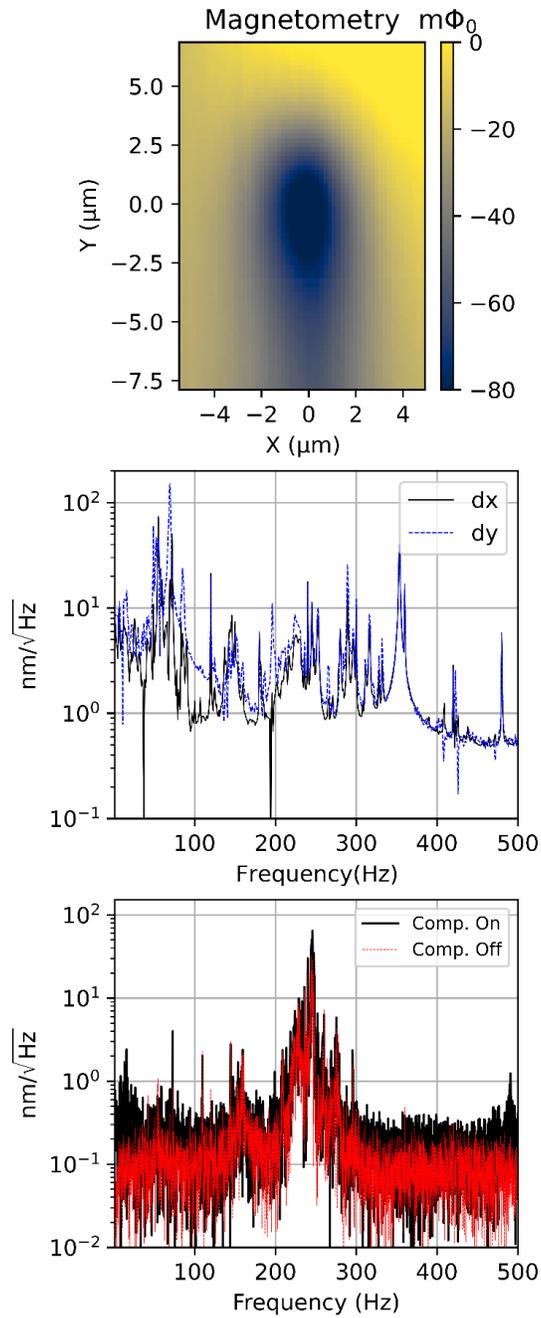

Figure 5: Extracting system vibrations using a vortex image and susceptibility measurements. (a) time averaged magnetometry image of a single vortex. (b) Vibrational spectral density extracted from the full data set. The fast scan direction is along the y-axis. (c) Vibration spectrum in the z-direction.

**Variable Sample Temperature**

The system can be configured with a thermally isolated sample holder allowing for variable sample temperature measurements. The thermal isolation was designed around experiments investigating Nb around $T_c$ = 9.2 K with the ability to quickly cool the sample for other measurements. A plot of the 4 K plate temperature as a function of sample temperature is shown in Figure 6(a), demonstrating the practicality of performing measurements with a sample temperature over 40 K. It takes around 3.5 minutes for the sample to get back to base temperature, allowing for rapid imaging at various temperatures.

To verify that the system and sensor can operate through the transition temperature of Nb, we gathered a series of susceptibility approach curves from just above $T_c$ to the base temperature, shown in Figure 6(b), on a 30 nm Nb film. We then fit the data at each temperature to a model of the expected response for a thin diamagnetic layer[38] and extracted a penetration depth for each temperature. When we fit these penetration depths to the expected trend with temperature[39], $\lambda(T) = \lambda(0)/\sqrt{1 - (T/T_c)^4}$, we extract a T = 0 K penetration depth of 122 nm ± 8 nm, consistent with other measurements of thin Nb films[40].

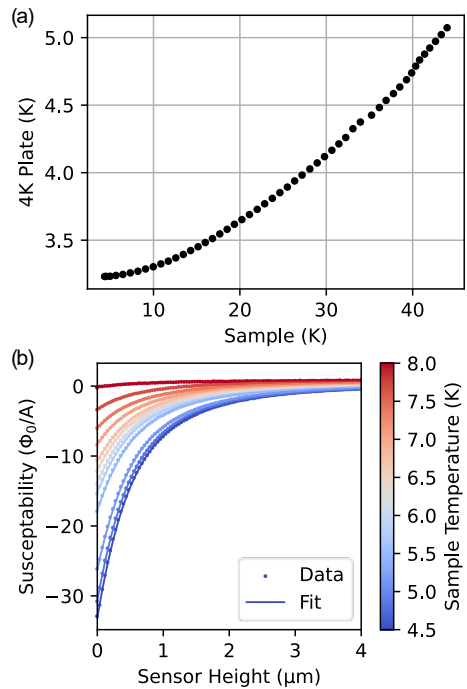

Figure 6: Variable sample temperature capability. (a) 4K stage temperature rise as the sample is heated. (b) Susceptibility approach curves from $T_c$ to base temperature.

While the data lead to a good measurement of the penetration depth, they also indicate Tc = 8.0 K for the film. Measurement of a 30 nm film using a commercial SQUID magnetometer indicate $T_c$ = 8.1 K, in good agreement with the value measured in the SSM. We also see a base temperature of 4.5 K which is 1.2 K higher than 3.3 K surface that the sample holder is mounted on. This is likely a result of insufficient thermalization of the leads for the heater and temperature sensor on the 4K stage.

**System Imaging**

Images are obtained using mechanically amplified piezoelectric scanners with a nominal range of 150 µm x 150 µm. The images shown below were captured with the y-axis piezo damaged, resulting in approximately half the nominal scan range. The system can operate in Earth's field (51.2 µT) as shown in Figure 7(a, c) which were captured by scanning over a 150 nm niobium film at 3.3 K. Here a thin field coil located under the sample was biased at 2 mA corresponding to an applied field of 53.4 µT. This applied field was used to counteract and then flip the field, resulting in vortices in the opposite direction, as indicated by the change in color from Figure 7 (a) to (c). With the shielding installed, the background field is reduced to 1 µT and the field can also be flipped and the vortex density increased as shown in Figure 7 (b, d). The "tail" observed in a single vortex is due to imperfect shielding in the SQUID sensor pickup area, and the small angle the sensor is mounted at to prevent the wire bonds from contacting the surface. The orientation and size of the "tail" is independent of fast scan direction and does not change with scan speed. This artifact can be corrected by post processing the data with an assumed point spread function, and other filtering can be performed to more precisely localize vortex positions[41].

The ability to do simultaneous magnetometry and susceptibility measurements is a powerful tool for navigating around superconducting circuits and structures and makes finding regions of interest considerably faster and easier. The images of a 150 nm continuous Nb film with a 15 µm grid of 5 µm lines on the right, Figure 8(c), and a 5 µm Nb meander biased at 100 µA, Figure 8(d), were obtained by passing 200 µA$_{RMS}$ at 1 kHz through the field coils on the sensor with the output from a lock in amplifier giving the resultant signal. In both images there is a strong signal from the 150 nm Nb and virtually no signal from the $SiO_2$ beneath. Magnetometry data can also be gathered synchronously by performing averaging on the raw output from the room temperature electronics. The scan speed is approximately 1 Hz so even a small amount of averaging removes the AC signal imparted from the field coils and the DC level represents the magnetometry. Figure 8(a) show the magnetometry signal taken at the same time as Figure 8(c). Here, many vorticities are visible and the fields penetrating the Nb grid are also imaged. Figure 8(b) shows the magnetometry signal from a 5 µm Nb wire current biased at 100 µA. The 5 µm wire is narrow enough that it is not energetically favorable to have vortices penetrate it[42].

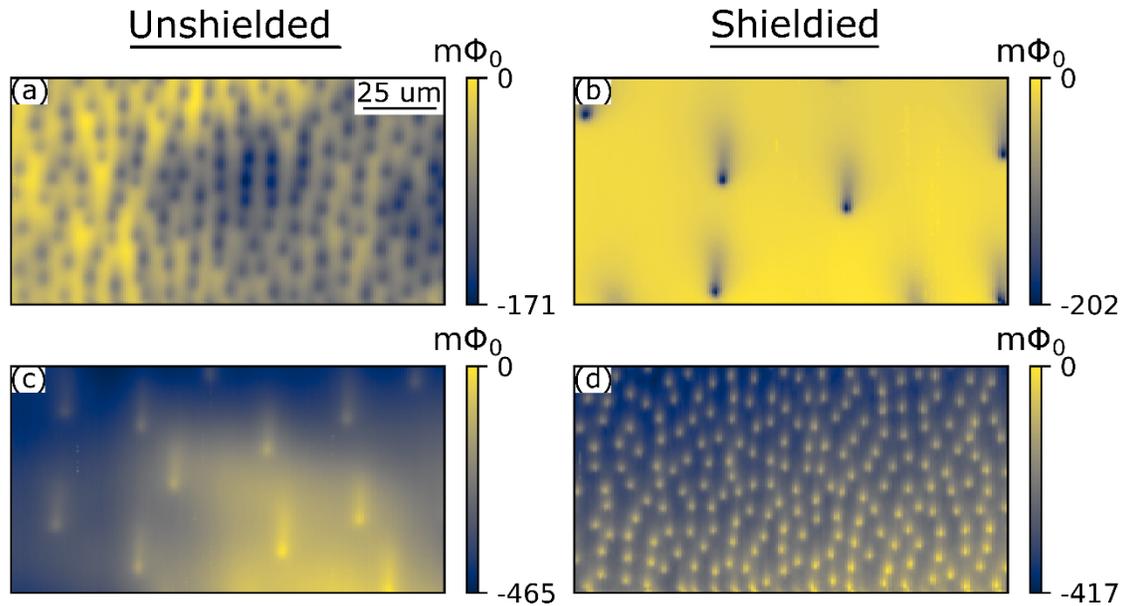

Figure 7: Images of vortices in a 150 nm Nb film with the system unshielded and shielded. (a) magnetometry image of vortices in earths field. (b) magnetic shielding installed resulting in a 1 µT background field and lower vortex density. (c) Same location and conditions as (a) but with field coil on. (d) Field coil used to increase vortex density in the same location and shielding as (b). The scale bar from (a) applies to all of the images.

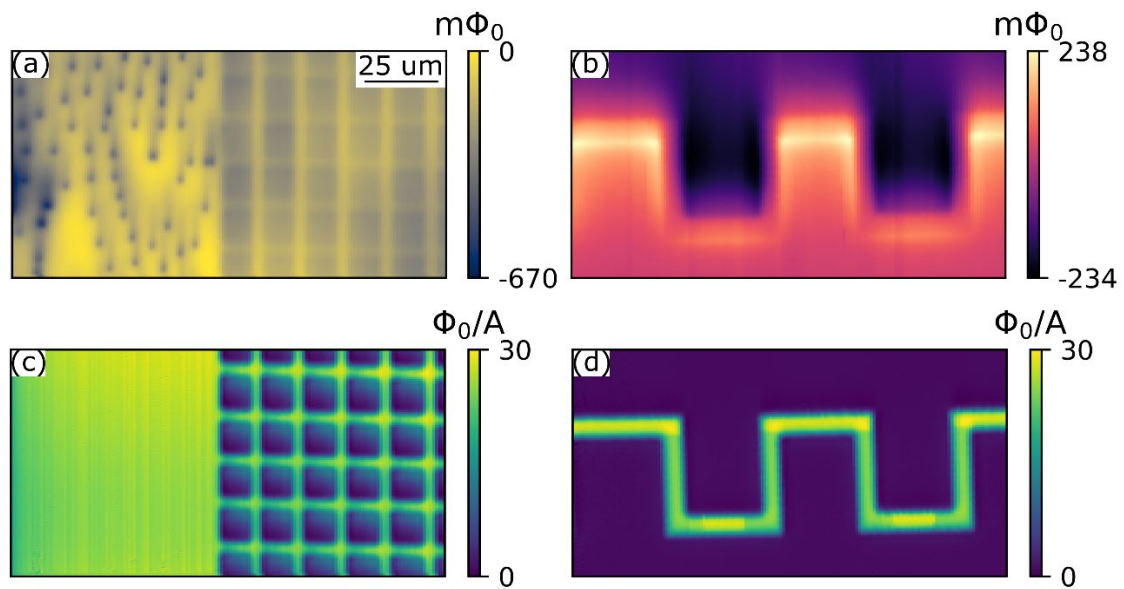

Figure 8: Images of a 150 nm continuous Nb film with 15 µm grid on the right side of the image (a and c), and a 5 um Nb meander biased with 100 µA (b and c). The magnetometry (a and b) and susceptibility (c and d) measurements were taken simultaneously for the two samples. Note that the color map for (c) was chosen to differentiate the fact the fields from currents are being shown. The scale bar from (a) applies to all of the images.

**CONCLUSION**

We have presented a cryogen-free, variable-temperature, scanning SQUID microscope featuring up to forty, 20 GHz bandwidth electrical lines. RF signals can be brought to a DUT using a combination of a PCB connected to a non-magnetic, cryogenic, chip socket and a superconducting interposer which is indium bump-bonded to the DUT and has an aperture allowing access for the scanned SQUID sensor. This arrangement has low electrical losses up to 20 GHz and it shows promise as a solution for measuring electrical margins in superconducting digital logic circuits as well as flux/vortex imaging in these circuits in a single cooldown. The system also has variable sample temperature capabilities allowing for measurements of susceptibility through the transition temperature of technologically relevant superconductors.